\documentclass[runningheads]{llncs}
\usepackage[T1]{fontenc}
\usepackage{graphicx}
\usepackage{svg}
\usepackage{amssymb}
\usepackage{amsmath}
\usepackage{bbm}
\usepackage{multirow}
\usepackage{booktabs}
\usepackage{algorithm}
\usepackage{algpseudocode} 
\usepackage{wasysym}
\usepackage{cite}

\begin{document}
\title{Contextual Scalarisation Thompson Sampling \\for multi-objective decisions in public media}

\titlerunning{Contextual Scalarisation Thompson Sampling}
\author{Théo Maëtz\inst{1},
Luc Guillet\inst{1},
Andrea Cavallaro\inst{2}}
\authorrunning{Théo Maëtz, Luc Guillet, Andrea Cavallaro}
\institute{
Radio Télévision Suisse, Geneva, Switzerland\\
\email{theo.maetz@rts.ch}\\
\email{luc.guillet@rts.ch}
\and
EPFL, Lausanne, Switzerland\\
\email{andrea.cavallaro@epfl.ch}
}

\maketitle              % typeset the header of the contribution
% Springer Licence to Publish, section 4(b): Submitted Manuscript acknowledgement
{\renewcommand{\thefootnote}{}\footnotetext{This preprint has not undergone peer review (when applicable) or any post-submission improvements or corrections. The Version of Record of this contribution is published in Pattern Recognition: 28th International Conference, ICPR 2026, Lyon, France, August 17--22, 2026, Proceedings, Part XIII, and is available online at \doi{10.1007/978-3-032-31927-2_24}.}}
%The abstract should briefly summarise the contents of the paper in 150--250 words.
\begin{abstract}
Recommender systems may operate under multiple, competing objectives. For example, audience reach, cultural values, public service mandate, and operational constraints must be balanced in editorial decisions of public service media. Existing approaches relying on fixed combinations of objectives or Pareto-based optimisation do not adapt to changing priorities across situations. In this paper, we propose Contextual Scalarisation Thompson Sampler (CSTS), a multi-objective contextual bandit method that learns to weight objectives as a function of the observed context. We evaluate CSTS on real programming data from Radio Télévision Suisse, the Swiss national broadcaster, showing improved contextual relevance and better alignment with expert curation practices compared to fixed weight and standard contextual bandit approaches.
\end{abstract}
%

%%%%%%%%%%%%%%%%%%%%%%%%%%%
\section{Introduction}
\label{sec: intro}

Recommender systems that optimise solely for short-term engagement can narrow exposure to a small set of highly popular items, amplify popularity biases and reduce diversity in what audiences see~\cite{Nguyen_Hui_Harper_Terveen_Konstan_2014,Pariser_2012}. Rather than optimising a commercial objective such as engagement, public service media organisations are mandated to act in the public interest. Hence, their programming decisions must balance multiple, sometimes conflicting, criteria \cite{charte_rts}. Examples of such criteria include reaching a broad audience while promoting cultural values with local relevance, representing all voices including minorities, supporting local productions, respecting legal and significant budgetary constraints, and maintaining long-term viewer trust. The corresponding decisions, which need to be transparent, can be posed as a multi-objective recommendation problem, where each criterion is modelled through a corresponding value signal.

Public service media scheduling has traditionally been driven by expert curators using manual workflows. For each programming decision, curators visually inspect catalogues of films and shows, relying on their expertise to weigh several factors at once, depending on the situation.  In practice, much of the broadcasting grid is constrained by ingrained viewing habits and recurring formats (e.g.\ news, magazines), so the room for manoeuvre is limited in many time slots. By contrast, scheduled film slots are where curators typically have the most flexibility and where the editorial stakes are highest, which is why in this work we focus on {\em movie recommendations}. For each time slot curators identify a small set of plausible candidates and then make a final choice within this shortlist. Therefore, a recommender system should help at the slate level (i.e., the candidate list for a time slot) by surfacing a ranked list of films that largely overlaps with those an expert would consider for that specific context.  The overall programming grid (i.e., the full schedule across time slots) is still mostly curated by hand, with limited algorithmic support beyond basic filtering and search by genre, keywords, production year, award nominations, or synopsis. While these tools help narrow down thousands of titles to a manageable shortlist, they do not account for editorial tradeoffs nor adapt to different programming contexts. Scheduling systems such as the Mediagenix Platform \cite{Corrigan_2025} are widely used across European public service media for catalogue management and grid scheduling, and include automation modules for secondary events or promotional content. However, in the premium editorial slots we focus on, the final programme selection and tradeoffs remain largely driven by curators: current systems are mostly used to manually apply basic searches and filters, a process that does not scale well to large catalogues. 

To address this challenge, we designed a decision support system for curators to navigate the catalogue more easily by surfacing a shortlist of plausible options for the curator to make the final decision. To this end, we propose Contextual Scalarisation Thompson Sampler (CSTS), a multi-objective recommender formulated as a contextual multi-armed bandit. CSTS models utility as a learned, context-dependent scalarisation of a vector of value signals. Unlike prior work that relies on fixed weights~\cite{Riabchuk_Hagel_Germaine_Zharova_2024,Rodriguez_Posse_Zhang_2012,qassimi_multi-objective_2025,jannach_survey_2023} or static Pareto frontiers~\cite{ribeiro_pareto-efficient_2012,jin_pareto-based_2024}, CSTS decomposes recommendation quality into multiple bounded value signals and learns a context-dependent weight vector to explore tradeoffs between objectives, while retaining the interpretability and data efficiency of scalar reward bandits. The exploration happens via Thompson sampling~\cite{Russo_Roy_Kazerouni_Osband_Wen_2020,agrawal_thompson_2014}. 
\begin{figure}[t]
\begin{center}

\includegraphics[width=1\textwidth]{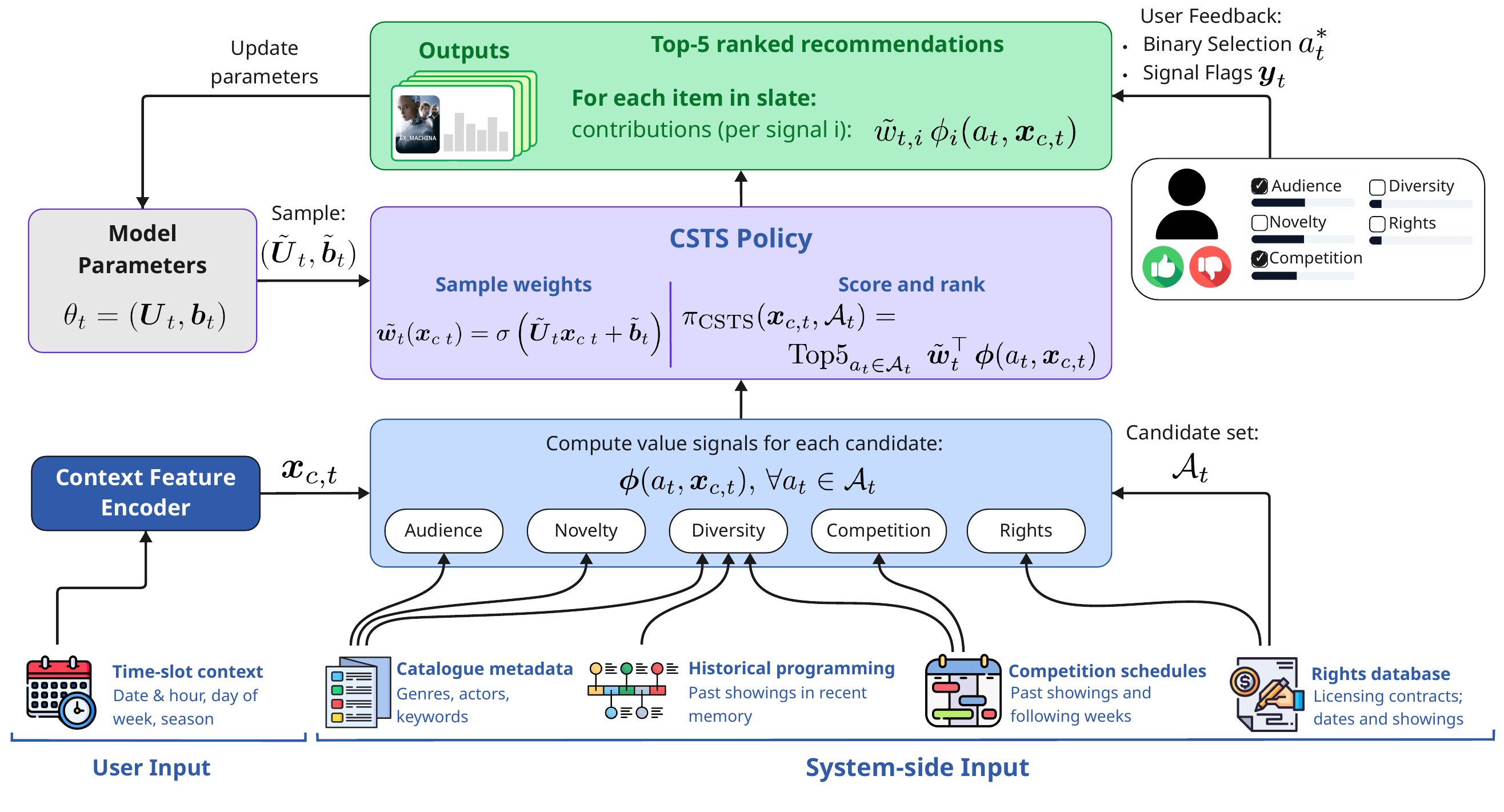}
\end{center}
\caption{Decision-support with the Contextual Scalarisation Thompson Sampler (CSTS). Given a time-slot context $\boldsymbol{x}_{c,t}$ (constructed from broadcast slot descriptors), system-side data such as catalogue metadata, historical programming, competition schedules, and the rights database, CSTS determines the available candidate set $\mathcal{A}_t$ and computes five value signals $\boldsymbol{\phi}(a_t,\boldsymbol{x}_{c,t})$, namely audience, novelty, diversity, competition, and rights, for each candidate title $a_t\in\mathcal{A}_t$. CSTS samples model parameters $\tilde{\theta}_t=(\tilde{U}_t,\tilde{b}_t)$ to produce context-dependent weights $\tilde{\boldsymbol{w}}_t(\boldsymbol{x}_{c,t})=\sigma(\tilde{U}_t\boldsymbol{x}_{c,t}+\tilde{b}_t)$ and scores each candidate by $\tilde{u}_t(a_t)=\tilde{\boldsymbol{w}}_t^\top \boldsymbol{\phi}(a_t,\boldsymbol{x}_{c,t})$. The Top-5 scored items form the ranked slate. Each recommendation is accompanied by per-signal contributions $\tilde{w}_{t,i}\phi_i(a,\boldsymbol{x}_{c,t})$. The curator feedback consisting of the binary selection and optional value signal flagging $\boldsymbol{y}_t$, is used to update $\theta_t$.   } 
\label{fig: full_system}
\end{figure}

%%%%%%%%%%%%%%%%%%%%%%%%%%%%%%%%%%%%%%%%%%%%%%
\section{Related Works}

A {\em bandit} problem is a sequential decision-making game between a learner and an environment~\cite{Lattimore_Szepesvári_2020}. At each round the learner chooses an action from a fixed set, observes only the reward for that action, and uses this feedback to improve future choices. A bandit algorithm specifies how the learner selects actions based on past interactions. Thompson sampling is a bandit algorithm that maintains a distribution over reward models and, at each decision step, samples model parameters from this distribution and implements the optimal action under the sample~\cite{Russo_Roy_Kazerouni_Osband_Wen_2020}.  Thompson sampling was adopted in recommender systems for  online matrix-factorisation to drive exploration in large catalogues~\cite{Kawale_Bui_Kveton_Tran-Thanh_Chawla_2015}. 

{\em Contextual bandits} let the learner observe a context describing the current situation (e.g.~user or time features) before choosing an action~\cite{li_contextual-bandit_2010}. In this regard the quality of a recommendation is not fixed but modelled as context-dependent, because the same item may be relevant for a situation (or user) and irrelevant in another~\cite{Adomavicius_Mobasher_Ricci_Tuzhilin_2011}. Linear contextual bandits with a scalar reward improve per-user article choice (click-through rates)~\cite{li_contextual-bandit_2010}. Change detection can be combined with disjoint and hybrid payoff models, where each arm has its own piecewise-stationary preference vector, so different items can change at different times~\cite{xu_contextual-bandit_2020}. The reward depends both on an arm-specific preference vector and on a global coefficient vector shared by all arms, so part of user preference can drift jointly across items while part changes separately for each arm. A hypernetwork may map contextual features to the parameters of the contextual bandit model, still targeting a single scalar measure of user engagement~\cite{shen_hyperbandit_2023}. Explanations layered on top of an underlying single-objective bandit inside the reward can augment contextual bandits for music playlist recommendations~\cite{mcinerney_explore_2018}. Context-specific importance weights can be combined with context-independent user preferences over reviews to produce a single rating for each item in a given context~\cite{Chen_Chen_2015}. Thompson sampling can be extended to contextual bandits with linear payoffs~\cite{agrawal_thompson_2014}. A scalable neural contextual bandit with an epistemic neural network architecture that represents uncertainty in a compact way implements Thompson sampling with few neural forward passes~\cite{zhu_scalable_2023}. However, this work still optimises a single scalar behavioural reward, such as clicks or user ratings.

{\em Multi-objective recommender systems} go beyond accuracy and may consider also diversity, novelty and coverage~\cite{jannach_survey_2023}. Fixed scalarisation with predefined weights  collapses all metrics into a single utility score~\cite{Riabchuk_Hagel_Germaine_Zharova_2024,jannach_survey_2023,Rodriguez_Posse_Zhang_2012}, for example by linearly combining accuracy and business metrics~\cite{Rodriguez_Posse_Zhang_2012}. Pareto-based methods, typically designed as offline systems, treat recommendations as genuine multi-objective optimisation and  provide a list of optimal solutions to choose from (or assume a static policy choice about where to operate on the Pareto frontier). A Pareto-efficient strategy can generate recommendation lists that are non-dominated with respect to accuracy, diversity, and novelty~\cite{ribeiro_pareto-efficient_2012}. Approximate Pareto frontiers can be manually adjusted to tradeoff competing objectives for the user to select their preferred point~\cite{jin_pareto-based_2024}. 
Closer to our setting, Qassimi and Rakrak~\cite{qassimi_multi-objective_2025} study multi-objective contextual bandits for smart tourism recommendation, where each arm is associated with several utility components (e.g. user satisfaction, provider revenue), and the algorithm balances them with prespecified scalarisation of objectives. 
The tradeoff between objectives is usually fixed at an a priori belief or tuned at a global level, rather than adapted to each decision context.

Table~\ref{tab:related_works} summarises the main features of bandit and multi-objective recommenders.
\begin{table}[t!]
    \centering
     \caption{Comparison of bandit and multi-objective recommenders. Key -- {\bf Ref.}: reference, {\bf B}: indicates whether it is formulated as a Bandit algorithm,  {\bf M}: specifies whether Multiple objectives (value signals) are modelled, {\bf C}:  indicates whether the method learns Context-dependent weightings over objectives, {\bf P}: marks methods that perform Pareto-based selection or optimisation, and {\bf E}:  marks methods that are Explainable.}
    \begin{tabular}{c@{\hskip 0.1in}l@{\hskip 0.2in}c@{\hskip 0.2in}c@{\hskip 0.2in}c@{\hskip 0.2in}c@{\hskip 0.2in}c@{\hskip 0.3in}l}
    \hline
       {\bf Ref.} & {\bf Method} & {\bf B} & {\bf M}   & {\bf C}  & {\bf P}  & {\bf E}   & {\bf Dataset(s)} \\
    \hline
        
        \cite{agrawal_thompson_2014}    & Linear TS bandit              & \CIRCLE & \Circle & \Circle & \Circle & \Circle & -- \\
        \cite{Kawale_Bui_Kveton_Tran-Thanh_Chawla_2015} 
                                        & Particle TS                   & \CIRCLE & \Circle & \Circle & \Circle & \Circle & -- \\
        \cite{zhu_scalable_2023}        & ENR                           & \CIRCLE & \Circle & \Circle & \Circle & \Circle & News, Social Media \\
        \cite{xu_contextual-bandit_2020}& Dynamic ctx. bandit           & \CIRCLE & \Circle & \Circle & \Circle & \Circle & News, Music \\
        \cite{shen_hyperbandit_2023}    & HyperBandit                   & \CIRCLE & \Circle & \Circle & \Circle & \Circle & News, Map POI \\
        \cite{li_contextual-bandit_2010} & LinUCB                        & \CIRCLE & \Circle & \CIRCLE & \Circle & \Circle & News \\\cite{mcinerney_explore_2018}   & Bart                          & \CIRCLE & \Circle & \Circle & \Circle & \CIRCLE & Music \\
        \cite{Chen_Chen_2015}           & Ctx. opinions Rec.            & \Circle & \Circle & \CIRCLE & \Circle & \Circle & Reviews \\
        \cite{Riabchuk_Hagel_Germaine_Zharova_2024} 
                                        & Utility-based scheduler       & \Circle & \Circle & \Circle & \Circle & \Circle & Energy consumption \\
        \cite{Rodriguez_Posse_Zhang_2012} 
                                        & Global weighted sum           & \Circle & \CIRCLE & \Circle & \Circle & \Circle & LinkedIn \\
        \cite{ribeiro_pareto-efficient_2012} 
                                        & Pareto-efficient hybrid.      & \Circle & \CIRCLE & \Circle & \CIRCLE & \Circle & Movies, Music \\
        \cite{jin_pareto-based_2024}    & PMORS                         & \Circle & \CIRCLE & \Circle & \CIRCLE & \Circle & Short videos \\
        \cite{qassimi_multi-objective_2025} 
                                        & MOC-MAB                       & \CIRCLE & \CIRCLE & \Circle & \Circle & \CIRCLE & Tourism, TripAdvisor \\
        ours                              & CSTS                 & \CIRCLE & \CIRCLE & \textbf{\CIRCLE} & \Circle & \textbf{\CIRCLE} & TV programming \\
    \hline
    \end{tabular}
    \label{tab:related_works}
\end{table}
In our work, we assume a contextual bandit framework but consider multiple value signals for each action. Rather than optimising several objectives separately or approximating a Pareto frontier, we introduce a {\em contextual scalarisation mechanism} that combines these signals into a single reward based on the current context. This preserves the algorithmic simplicity and data efficiency of scalar reward contextual bandits while allowing the adaptation of the relative importance of objectives across situations.

%%%%%%%%%%%%%%%%%%%%%%%%%%%%
\section{Proposed approach}

Let $\mathcal{A}_t$ be the set of possible actions, namely choices of available titles, which vary over $t$ due to contractual constraints. Let $c$ be the context index and $\boldsymbol{x}_{c,t} \in \mathcal{X}$ the context vector that represents features describing the slot, channel, seasonality/time of year  at time $t$ in the context space, $\mathcal{X}\subseteq \mathbb{R}^p$, where  $p \in \mathbb{N}$ is the context dimension.  

Let a vector of objectives, $\boldsymbol{\phi}(a_t, \boldsymbol{x}_{c,t})$, encode how well action $a_t$ performs along each of the $N$ underlying objectives:
\begin{equation}
\boldsymbol{\phi}(a_t, \boldsymbol{x}_{c,t}) = \left(\phi_1(a_t, \boldsymbol{x}_{c,t}), \dots, \phi_N(a_t, \boldsymbol{x}_{c,t})\right),
\label{eq: value_signal}
\end{equation}
where $\boldsymbol{\phi}_i(a_t,\boldsymbol{x}_{c,t})\in [0,1]$ represents a specific objective, the $i$-th value signal for action $a_t$ in context $\boldsymbol{x}_{c,t}$.
In our programming instantiation, we define {$N=5$} objectives measuring (i) audience potential, (ii)  diversity, (iii)  novelty, (iv) competition with concurrently airing programmes, and (v) rights/contracts management. The relative contribution of the value signals may vary with $\boldsymbol{x}_{c,t}$. 

After selecting an action $a_t \in \mathcal{A}_t$ in context $\boldsymbol{x}_{c,t}$, the agent observes a binary reward $r_t \in \{0,1\}$. Let $\pi$ be a policy that  selects an action
\begin{equation}
a_t = \pi(\boldsymbol{x}_{c,t}, \mathcal{A}_t)
\end{equation}
for each context, $\boldsymbol{x}_{c,t}$, and candidate set, $\mathcal{A}_t$. We aim to learn $\pi$ that, on average over the different possible contexts, selects actions with maximal expected reward according to the contextual weighting of the value signals:

\begin{equation}
\mu_\pi(\boldsymbol{x}_{c, t}, \mathcal{A}_t)
=\mathbb{E}\!\left[r_t \mid \boldsymbol{x}_{c,t},\ \mathcal{A}_t,\ a_t=\pi(\boldsymbol{x}_{c, t}, \mathcal{A}_t)\right].
\end{equation}

This problem can be formalised as a multi-objective contextual multi-armed bandit problem. The idea is to exploit $\boldsymbol{\phi}(a_t, \boldsymbol{x}_{c,t})$ while learning from single bandit reward $r_t$ observed for the chosen action. 
For each $(a_t, \boldsymbol{x}_{c,t})$ the bounded and normalised value signal vector $\boldsymbol{\phi}(a_t, \boldsymbol{x}_{c,t}) \in [0,1]^N$ summarises how $a_t$ performs on the $N$ underlying objectives. 

In our specific use case, we adopt external contextual information such as the broadcast schedules of competing channels over the same period, the catalogue of all films that were available, additional movie metadata from TMDb \cite{tmdb} (e.g.~genres, production year), and the proximity of rights expiry in the catalogue, to compute the value signal vector $\boldsymbol{\phi}(a_t,\boldsymbol{x}_{c,t})$ for each candidate film $a_t$. For every programming decision, we reconstruct the complete context at time $t$ as a feature vector $\boldsymbol{x}_{c,t}$ (time of day, day of the week, public holidays, channel and competition schedule), the set of available movies $\mathcal{A}_t$ given rights and contractual constraints, the associated value signal vector $\boldsymbol{\phi}(a_t,\boldsymbol{x}_{c,t})$ for each $a_t \in \mathcal{A}_t$, and the item $a^\star_t$ actually selected by the curator. These components reflect the environment of the curator responsible for the programming decisions.

Rather than fixing a global weighting of these objectives as in \cite{Rodriguez_Posse_Zhang_2012,jannach_survey_2023,qassimi_multi-objective_2025,Riabchuk_Hagel_Germaine_Zharova_2024}, we let their relative importance depend on the context through a learned weighting function $\boldsymbol{w}: \mathcal{X} \rightarrow  \Delta^{N-1}$, with $\Delta^{N-1}$ being the probability simplex where all entries are non-negative and summing to one. We define a scalar utility, $u(a_t, \boldsymbol{x}_{c,t}) \in [0,1]$, as
\begin{equation}
    u(a_t, \boldsymbol{x}_{c,t}) = \boldsymbol{w}(\boldsymbol{x}_{c,t})^\top \boldsymbol{\phi}(a_t, \boldsymbol{x}_{c,t}).
\label{eq: scalar_utility}
\end{equation}

Given $u(a_t, \boldsymbol{x}_{c,t})$ and $\boldsymbol{\phi}(a_t, \boldsymbol{x}_{c,t})$, we learn a context-dependent weight function $\boldsymbol{w}(\boldsymbol{x}_{c,t})$ to tradeoff different objectives in each situation.
Analogous to applying a multinomial logistic layer over objectives \cite{Bishop_2006}, we parametrise this weighting function with a linear gating model over the context features: 
\begin{equation}
    \boldsymbol{w}_\theta(\boldsymbol{x}_{c,t}) = \mathrm{softmax}\!\left(\boldsymbol{U} \boldsymbol{x}_{c,t} + \boldsymbol{b}\right),
    \label{eq: context_gating}
\end{equation}
where $\theta = (\boldsymbol{U},\, \boldsymbol{b})$ are the model parameters:  $\boldsymbol{U} \in  \mathbb{R}^{N \times p}$ maps the context vector $\boldsymbol{x}_{c,t}$ to one logit per value signal and $\boldsymbol{b} \in  \mathbb{R}^{N}$ adds a baseline logit per signal, independent of context. To ensure that all entries are non-negative and sum to one, the summed logits are then passed through the softmax function $\mathbb{R}^N \rightarrow [0, 1]^N$. This keeps each objective visible through $\boldsymbol{\phi}(.,.)$, while expressing the tradeoff between them with a probability vector over objectives.

To balance exploration and exploitation when generating $\boldsymbol{w}(\boldsymbol{x}_{c,t})$, we use the contextual Thompson sampling framework~\cite{agrawal_thompson_2014} on $\theta = (\boldsymbol{U},\, \boldsymbol{b})$. 
\begin{algorithm}[t]
\caption{Contextual Scalarisation Thompson Sampling (CSTS) decision}
\label{alg:csts}
\begin{algorithmic}[1]
    \Require $\boldsymbol{x}_{c,t}$, $\mathcal{A}_t$, $\theta = (\boldsymbol{U}, \boldsymbol{b})$
    \State Sample $\tilde{\theta}_t = (\tilde{\boldsymbol{U}}_t,\tilde{\boldsymbol{b}}_t)$
    \State $\boldsymbol{\tilde{w}}_t \leftarrow \boldsymbol{w}_{\tilde{\theta}_t}(\boldsymbol{x}_{c,t})$ \Comment{Eq.~\eqref{eq: context_gating}}
    \For{each $a_t \in \mathcal{A}_t$}
        \State Compute $\boldsymbol{\phi}(a_t,\boldsymbol{x}_{c,t})$
        \State $\tilde{u}_t(a_t) \leftarrow \boldsymbol{\tilde{w}}_t^\top \boldsymbol{\phi}(a_t,\boldsymbol{x}_{c,t})$
    \EndFor
    \State $a_t^* \gets \arg\max_{a_t \in \mathcal{A}_t} \tilde{u}_t(a_t)$ \Comment{or select the top-$K$ actions}
    \State Observe reward $r_t$ for $a_t^*$
    \State $u_t(a_t^*) \leftarrow \boldsymbol{w}_\theta(\boldsymbol{x}_{c,t})^\top \boldsymbol{\phi}(a_t^*,\boldsymbol{x}_{c,t})$
    \State $\hat{p}_t \leftarrow \sigma\!\big(u_t(a_t^*)\big)$
    \State $\theta_{t+1} \leftarrow \textsc{UpdateParameters}(\theta_t, \boldsymbol{x}_{c,t}, r_t, \hat{p}_t)$
\end{algorithmic}
\end{algorithm}
Because the randomness enters at the level of $\theta$, the algorithm explores different plausible tradeoffs between objectives rather than injecting arbitrary noise in item scores. In parts of the context space where we have only few observed decisions and rewards for similar contexts, the model remains uncertain about the right weights. This uncertainty leads to larger variations (i.e.~more exploration) in the sampled weights $\boldsymbol{\tilde{w}}_t$. As we gather more feedback in well-covered regions, the sampled weights stabilise and the policy exploits what it has learned.
These components define the contextual bandit policy through Thompson sampling over $\theta$. At each step $t$, we draw $\tilde{\theta}_t$ from the current parameter distribution, compute the sampled
weights $\tilde{\boldsymbol{w}}_t = \boldsymbol{w}_{\tilde{\theta}_t}(\boldsymbol{x}_{c,t})$, and select
\begin{equation}
\pi_{\text{CSTS}}(\boldsymbol{x}_{c,t}, \mathcal{A}_t) \;=\; \arg\max_{a_t \in \mathcal{A}_t} \;\tilde{\boldsymbol{w}}_t^\top \boldsymbol{\phi}(a_t, \boldsymbol{x}_{c,t}).
\end{equation}

When recommending a ranked shortlist (slate) of size $K$, we return the Top-$K$ actions in $\mathcal{A}_t$ according to the same sampled utility.

After the curator selects an item, we observe the binary reward $r_t$ and update the parameters $\theta$ by comparing this outcome to the scalar utility assigned to the chosen action under the current weights:
\begin{equation}
    u(a_t, \boldsymbol{x}_{c,t})
    = \boldsymbol{w}_\theta(\boldsymbol{x}_{c,t})^\top
      \boldsymbol{\phi}(a_t, \boldsymbol{x}_{c,t}).
\end{equation}

We convert this scalar utility to a predicted acceptance probability with a sigmoid function, $\hat{p}_t = \sigma\!\big(u(a_t, \boldsymbol{x}_{c,t})\big)$, and update $\theta$ by minimising a logistic loss:
\begin{equation}
    \mathcal{L}_t(\theta)
    = -\Big[
        r_t \log \hat{p}_t
        + (1 - r_t) \log\big(1 - \hat{p}_t\big)
      \Big],
\end{equation}
using standard gradient descent \cite{Summa_Bottou_Goldfarb_Murtagh_Pardoux_Touati_2011}.

We maintain a simple parametric distribution over the parameters, which we use to draw $\tilde{\theta}_t$ in Algorithm~\ref{alg:csts}. This distribution is centred at the current parameters $\theta_t$, and its scale encodes the uncertainty about each parameter. We use a diagonal approximation, maintaining uncertainty per parameter via an RMS-type (Root Mean Square) accumulator of past gradients, as in RMSProp~\cite{tieleman2012rmsprop}. After each decision, the same gradient information used to update $\theta_t$ is also used to update these uncertainty terms.

Because the model scores actions through value signals $\boldsymbol{\phi}(a_t, \boldsymbol{x}_{c,t})$ and context-dependent weights $\boldsymbol{w}_\theta(\boldsymbol{x}_{c,t})$, for each objective we can explicitly see  the strength of the signal and the value of its weight in a given context. This allows the curator to optionally input through a guiding vector $\boldsymbol{y}_t \in \mathbb{R}^N$  which objectives are most relevant for a given decision, beyond just observing $r_t \in \{0,1\}$ for the chosen action.
We then use $\boldsymbol{y}_t$ to steer the learned weights $\boldsymbol{w}_\theta(\boldsymbol{x}_{c,t})$ for that specific situation through an auxiliary signal matching loss:
\begin{equation}
    \mathcal{L}_t^{\text{signal}}(\theta)
    = \frac{1}{2}\big\|
        \boldsymbol{w}_\theta(\boldsymbol{x}_{c,t})
        - {\boldsymbol{y}}_t
      \big\|_2^2.
\end{equation}

We add this loss to the bandit reward logistic loss for updating $\theta$ to turn curator feedback into an explicit signal about editorial priorities in a given context.

In the next section, we assess the performance of the proposed contextual multi-objective bandit, CSTS, in reproducing expert programming decisions and in terms of tradeoffs between value signals across different contexts.

%%%%%%%%%%%%%%%%%%%%%%%%%%%%%%%%%%%%%%%%%%%%%
\section{Evaluation}

%%%%%%%%%%%%%%%%%%%%%%%%%%%%%%%%%%%%%5
\subsection{Experimental setup}

 We compare CSTS with (i) static global weights, (ii) audience potential maximisation using a supervised audience rating regressor trained offline on past broadcasts, (iii) vanilla Thompson sampling \cite{Russo_Roy_Kazerouni_Osband_Wen_2020} and (iv) LinUCB contextual bandit \cite{li_contextual-bandit_2010}.
 The {\em static global weights} approach has a fixed global weight vector $\boldsymbol{w}_0 \in \Delta^{N-1}$. This is a non-contextual multi-objective baseline defined through curator discussions and scores candidates by $u(a_t,\boldsymbol{x}_{c,t}) = \boldsymbol{w}_0^\top \boldsymbol{\phi}(a_t,\boldsymbol{x}_{c,t})$. This
baseline tests how far one can go with a single, context-independent tradeoff between value signals. The static weights account for domain knowledge about typical tradeoffs across the grid. The {\em audience potential maximisation} approach mimics a standard recommender driven by engagement metrics. This  baseline uses only the audience potential value signal.
The {\em vanilla Thompson sampling} is a non-contextual multi-armed bandit baseline that uses the value signals $\boldsymbol{\phi}(a_t,\boldsymbol{x}_{c,t})$ but assigns them global, context-independent weights. We apply Thompson sampling directly on these weights, so that each value signal has a single posterior weight vector shared across all contexts. Finally, the {\em LinUCB contextual bandit} approach is a linear contextual bandit based on LinUCB \cite{li_contextual-bandit_2010}. 
For each candidate $a_t \in \mathcal{A}_t$ in context $\boldsymbol{x}_{c,t}$, we build a feature vector $\boldsymbol{\psi}(a_t, \boldsymbol{x}_{c,t})$ by concatenating the slot context and the value signals, i.e.\ $\boldsymbol{\psi}(a_t,\boldsymbol{x}_{c,t}) = [\,\boldsymbol{x}_{c,t}; \boldsymbol{\phi}(a_t,\boldsymbol{x}_{c,t})\,]$. The expected bandit reward $r_t \in \{0,1\}$ is modelled as a linear function $\mathbb{E}[r_t \mid a_t,\boldsymbol{x}_{c,t}] \approx \theta^\top \boldsymbol{\psi}(a_t,\boldsymbol{x}_{c,t})$. LinUCB selects actions using this prediction and an upper confidence bonus that favours candidates with more uncertain rewards. In contrast to our contextual Thompson bandit, which first maps context to weights over value signals and samples these weights with uncertainty, LinUCB treats context and signals as a flat feature vector, optimises a single scalar reward, and explores through an explicit confidence bonus rather than parameter sampling.

We use two years of historical programming logs, where each record corresponds to a curator choosing a film for a specific RTS time slot. We evaluate all recommenders under comparable conditions by adopting an offline replay protocol on the historical logs. We iterate over the reconstructed programming decisions in chronological order and, at each time $t$, let each policy rank the corresponding candidate set $\mathcal{A}_t$ given its context. We then take the policy's top-$K$ recommendations for comparison with the logged choice $a^\star_t$.

\subsection{Metrics}

Slate-ranking metrics quantify how many relevant options the system retrieves for a slot and how prominently it presents them. In our setting, relevance is defined from RTS editorial practice; derived from slot-specific relevance rules provided by RTS TV curators (e.g.\ genre, tone, audience targeting). We use two standard slate-based ranking metrics, Hit@K and NDCG@K, for Top-$K$ recommendation and slate evaluation \cite{He_Liao_Zhang_Nie_Hu_Chua_2017,Zhou_Shen_Guo_Wu_Ma_2025}.
Let $\mathrm{TopK}_t$ denote the top-$K$ slate returned by a policy at time $t$, let $R_t \subseteq \mathcal{A}_t$ be the set of items deemed relevant for that slot, and let $T$ denote the number of programming decisions in the evaluation set. Hit@K measures how often at least one relevant item appears in the top-$K$:
\begin{equation}
\mathrm{Hit@K} = \frac{1}{T} \sum_{t=1}^T
\mathbbm{1}\{ R_t \cap \mathrm{TopK}_t \neq \emptyset \}, 
\end{equation}
where $\mathbbm{1}\{.\}$ is the indicator function, equal to $1$ if its condition holds and $0$ otherwise.

The normalised discounted cumulative gain at cutoff $K$, NDCG@K, accounts for the positions of relevant items within the slate. Let $\mathrm{rel}_t(a_t) \in \{0,1\}$ indicate whether item $a_t \in \mathcal{A}_t$ is
relevant at time $t$, and define
\begin{equation}
\mathrm{DCG}_t@K = \sum_{i=1}^K
\frac{\mathrm{rel}_t(a_{t,i})}{\log_2(1+i)}, 
\end{equation}
and
\begin{equation}
\mathrm{NDCG@K} = \frac{1}{T} \sum_{t=1}^T
\frac{\mathrm{DCG}_t@K}{\mathrm{IDCG}_t@K},
\end{equation}
where $\mathrm{IDCG}_t@K$ is the DCG of an ideal slate at time $t$, obtained by placing all items with $\mathrm{rel}_t(a_t)=1$ in the top positions (up to $K$).

We report both {\em strict} and {\em relaxed} relevance. In strict relevance, $R_t=\{a_t^\star\}$ where $a_t^\star$ is the historically scheduled film for slot $t$. In relaxed relevance, $R_t$ contains all films in $\mathcal{A}_t$ that satisfy the curator-provided slot criteria for time $t$ (e.g.\ Saturday family, Wednesday "classics", Friday youth-oriented). This captures both alignment with the logged choice and the ability to retrieve a slate coherent with the editorial intent of the context.

\subsection{Ranking Performance and Tradeoff Analysis}

We first assess contextual appropriateness through ranking metrics and then  how policies balance the underlying value signals. 

Table~\ref{tab:ranking_results} compares 75 test decisions across four key time slots. CSTS achieves very high contextual relevance through its adaptive scalarisation. On strict matching, fixed weights achieve the highest Hit@10 (20.0\%). This is not entirely surprising as we cannot fully observe all the considerations the curator had at decision time which led to that exact selection, and many programming choices are inherently punctual (e.g. reacting to competing events or last-minute constraints) and thus only partially captured in our context features. A fixed global weighting defined by the curators can therefore align more closely with the exact historical picks. CSTS takes a different approach by looking for films that best satisfy the situation at hand. The goal is not to mimic historical choices (which may be suboptimal or inconsistent), but to identify films that satisfy context-dependent criteria. Across both strict and relaxed metrics, CSTS consistently outperforms LinUCB \cite{li_contextual-bandit_2010}, indicating that context-dependent scalarisation brings benefits beyond linear contextual scoring alone.

\begin{table}[t]
\centering
\caption{Ranking performance on contextual relevance at $K=10$ (expressed as Hit@10 and NDCG@10) for $N=75$ programming decisions. CSTS achieves the highest relaxed contextual relevance, while remaining competitive under strict matching.}
\begin{tabular}{lcccc}
\hline
& \multicolumn{2}{c}{{\bf Strict}} & \multicolumn{2}{c}{{\bf Relaxed}} \\
{\bf Policy} {\hskip 0.2in}  & Hit@10 {\hskip 0.2in} & NDCG@10 {\hskip 0.2in}& Hit@10 {\hskip 0.2in}& NDCG@10 {\hskip 0.2in}\\
\hline
LinUCB        & 0.187          & 0.101          & 0.947          & 0.367          \\
Static weights& \textbf{0.200} & \textbf{0.108} & 0.920          & 0.336          \\
Audience max. & 0.027          & 0.017          & 0.813          & 0.258          \\
Vanilla TS    & 0.013          & 0.004          & 0.800          & 0.309          \\
CSTS           & 0.187          & 0.101          & \textbf{0.987} & \textbf{0.376} \\
\hline
\end{tabular}
\label{tab:ranking_results}
\end{table}

To isolate the {\em effect of contextual weights}, we compare the Contextual Thompson sampling with the context-independent weights from the vanilla Thompson sampling and the fixed global weights of the static recommender in Table~\ref{tab:ranking_results}. CSTS achieves 98.7\% contextual relevance, substantially outperforming vanilla Thompson (80.0\%, +18.7 percentage points (pp)) and Static (92.0\%, +6.7pp). The gap between CSTS and vanilla Thompson isolates the effect of context-adaptation: both use Thompson sampling and learn from the same data, but only CSTS adapts weights by context. CSTS's learned context-adaptive weights outperform hand-crafted global weights by 6.7pp. Interestingly, Static (92.0\%) outperforms vanilla Thompson (80.0\%) despite not learning, suggesting that domain expertise initialisation provides a strong baseline that global weight learning struggles to surpass without adaptation to context. This can be understood with vanilla Thompson having to learn a single set of global weights from sparse, noisy bandit feedback, under the assumption that the same tradeoff holds in every context. When the algorithm is forced to fit one global compromise, it averages over these conflicting patterns and washes out useful structure. 

\begin{figure}[t]
\begin{center}
\includegraphics[width=0.7\textwidth]{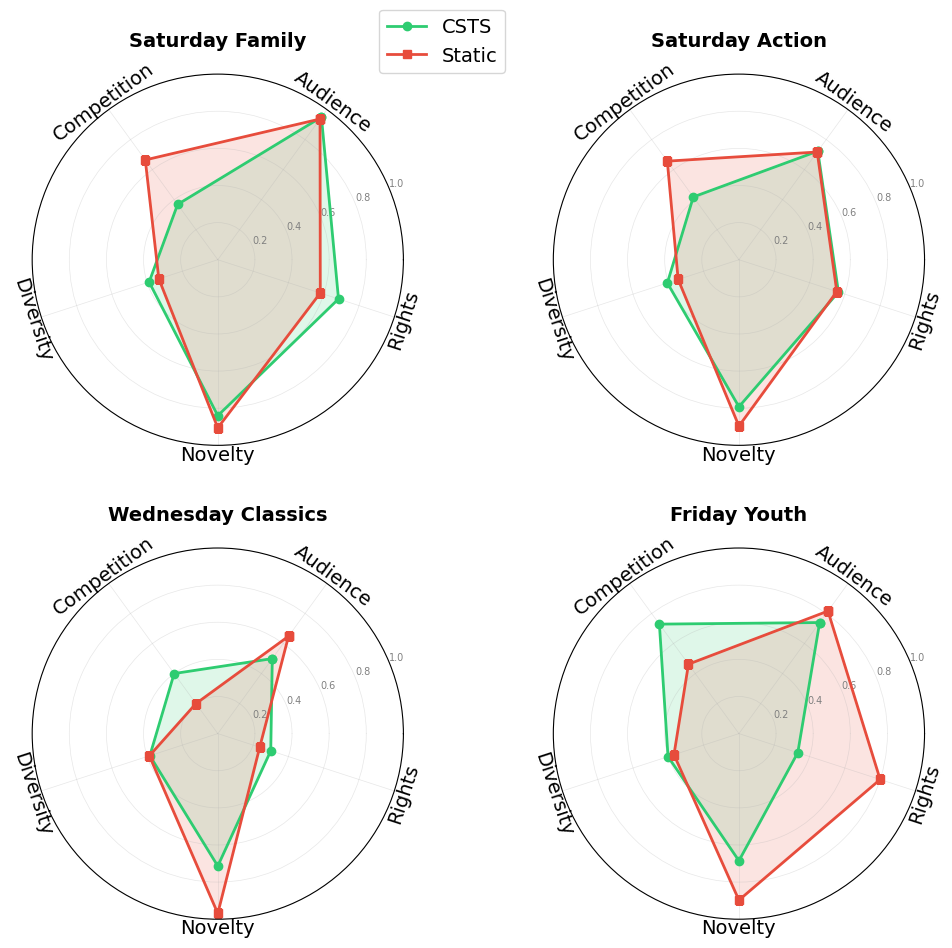}
\end{center}
\caption{Value signal profiles of top-1 recommendations for CSTS and Static across four programming contexts. CSTS aligns more frequently with Diversity and Rights urgency objectives (3 out of 4 contexts each), while Static maintains higher Novelty in every context.  Across contexts, the profiles differ by objective and neither method is uniformly higher across all value signals.
} 
\label{fig: radar_plot}
\end{figure}

Figure~\ref{fig: radar_plot} shows mean value signals for the top-1 recommendations across the four key time slots, comparing CSTS to fixed scalarisation. 
We can notice that CSTS and Static make different choices about what to prioritise. CSTS recommendations tend to have higher diversity (in 3 out of 4 contexts) and better address rights urgency (also 3 out of 4 contexts). CSTS learns from past curator decisions, leaning towards catalogue efficiency: making sure we use our available content well and manage rights effectively. On the other hand, static recommendations consistently favour novelty across all contexts.
Static uses weights defined by domain experts, which emphasises audience engagement through novelty and competitive positioning. Importantly, neither policy dominates across all objectives. The choice between them reflects organisational priorities.

\subsection{Ablation Study}

\begin{figure}[t!]
\includegraphics[width=\textwidth]{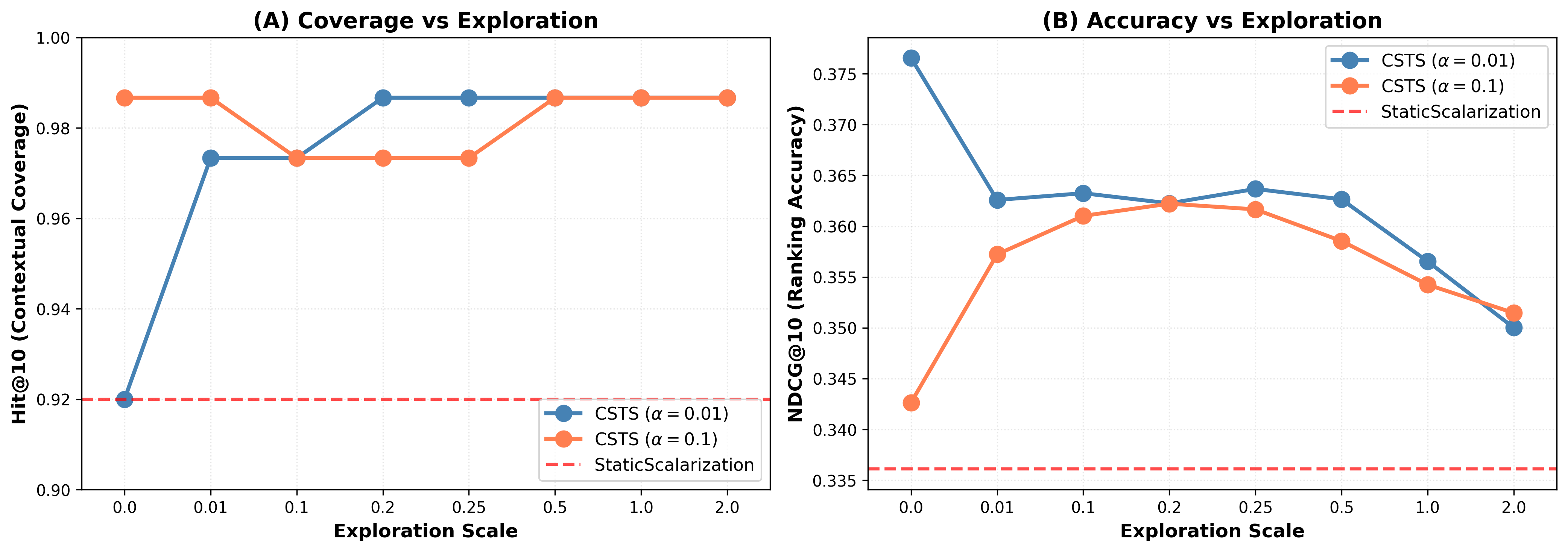}
\caption{Effect of exploration scale $\kappa$ on CSTS ranking performance under conservative ($\alpha = 0.01$) and aggressive ($\alpha = 0.1$) learning. Reporting relaxed contextual relevance metrics for CSTS, averaged over 75 test key time slots. Small $\kappa$ yields near-greedy exploitation of the learned utility, while large $\kappa$ encourages exploration through more diverse weight samples. Conservative learning yields stable performance where greedy already performs well, while more aggressive learning ($\alpha = 0.1$) benefits from moderate exploration (e.g. $\kappa = 0.2$) before performance degrades at very high $\kappa$.} 
\label{fig: exploration_graph}
\end{figure}

\noindent {\bf Effect of Exploration}.
Thompson sampling explores by sampling noisy parameters $\tilde{\theta}_t = (\tilde{U}_t, \tilde{b}_t)$. The noise level is controlled by  $\kappa$, the exploration scale: by varying from $\kappa=0$ (i.e. greedy exploitation, no sampling) to larger values (strong exploration), we quantify the effect of exploration (Figure~\ref{fig: exploration_graph}).
Greedy exploration achieves $\text{NDCG}@10=0.343$, which is below optimal.  With  a higher learning rate ($\alpha = 0.1$), greedy could be overfitting to training data; $\kappa \in [0.1,\,0.2]$ (optimal exploration) achieves the best accuracy ($\text{NDCG}@10=0.367$,  +7.0\% vs greedy) by regularising learning; $\kappa>1.0$ degrades performance by adding excessive noise (over-exploration). Coverage stays around 97–99\%, but dips slightly at the scales that give the best accuracy, highlighting an accuracy–coverage tradeoff. This contrasts with the conservative setting ($\alpha = 0.01$), where greedy attains the highest accuracy but suffers from lower coverage. Conservative learning prevents overfitting, making greedy exploitation effective at ranking accuracy within familiar contexts: with $\alpha = 0.01$, exploration mainly enables generalisation (trading accuracy for coverage), whereas with $\alpha = 0.1$, exploration prevents overfitting and thus improves accuracy, showing the need for calibrated exploration rather than purely greedy policies.

\noindent {\bf Effect of value signals}. In order to assess the importance of our multi-objective optimisation, we systematically remove each value signal and retrain CSTS. 
Table~\ref{tab:signal_ablation} shows ablation results. The performance drop in ranking metrics when removing each signal supports their contribution to contextual relevance.
\begin{table}[t!]
\centering
\caption{Ablation study of value signals in CSTS under relaxed contextual relevance. We report relaxed Hit@10 and NDCG@10 when using all five value signals and when removing each signal in turn. Results are averaged over the 75 test programming decisions. Removing either Competition or Audience causes the largest degradation in contextual relevance, while removing Diversity, Novelty, or Rights still harms performance to a lesser extent.}
\label{tab:signal_ablation}
\begin{tabular}{lcccc}
\hline
Configuration & Hit@10 & NDCG@10 & $\Delta$Hit@10  & $\Delta$NDCG@10 \\
\hline
Full (5 signals) & \textbf{0.987} & \textbf{0.376} & --  & -- \\
\hline
w/o Competition & 0.880 & 0.336 & -0.107 & -0.040 \\
w/o Audience & 0.893 & 0.310 & -0.093 & -0.066 \\
w/o Diversity & 0.920 & 0.372 & -0.067 & -0.004 \\
w/o Novelty & 0.973 & 0.373 & -0.013 & -0.004 \\
w/o Rights & 0.987 & 0.338 & -0.000 & -0.039 \\
\hline
\end{tabular}
\end{table}
Comparing to a single objective such as audience maximisation (81.3\% Hit@10, Table~\ref{tab:ranking_results}), even the weakest multi-objective configuration (w/o Competition: 88.0\%) outperforms pure audience maximisation (+6.7pp). 

\subsection{Limitations}

CSTS in its current form has two limitations. First, the value signals are hand-designed, scaled, and bounded. Their relative scales influence how differences on each dimension translate into changes in the scalar utility and learned weights. Second, the quality of these signals depends on external metadata and scheduling data (e.g.~catalogue, third-party metadata, competitor schedules), which are incomplete or noisy. Missing or incorrectly matched metadata can cause relevant titles to be misused or excluded from candidate sets, which is problematic in a public service media setting where unjustified exclusion risks being perceived as a form of bias or censorship.

%%%%%%%%%%%%%%%%%%%%%%%%%%%%%%%%%%%%%%%%%%%%%%
\section{Conclusion}

We proposed CSTS (Contextual Scalarisation Thompson Sampler), a decision support model based on a multi-objective contextual bandit framework with adaptive scalarisation for public service media programme scheduling. The framework is designed to support media curators in navigating large catalogues under multiple editorial objectives covering audience, diversity, novelty, competition, and contracts management. CSTS learns to weight these objectives based on the current broadcasting context using a Thompson-style randomised exploration strategy. Unlike fixed weight or Pareto-based approaches, CSTS dynamically adjusts tradeoffs, which leads to higher contextual relevance and better alignment with expert curation practices.

Experiments on two years of real programming data show a tradeoff across evaluation settings in terms of contextual relevance and editorial alignment. Under strict matching to historical choices, fixed scalarisation achieves the highest hit rate, reflecting that carefully tuned global editorial weights remain a strong reference point. However, CSTS improves relaxed contextual relevance, reaching near complete coverage of relevant sets across key time slots defined by the curators. 

As future work, we will move from hand-crafted value signals to a more systematic design process with simple rules and diagnostics to check that signals are on comparable scales and do not dominate the utility solely due to their own variance across the candidate set. Furthermore, we will investigate the integration of digital audience measurements alongside linear ratings.

\subsubsection{Acknowledgments} This work was supported by Radio Télévision Suisse. We thank Adèle Cserpes and Barbara Karkin for their continued feedback throughout this project.

\bibliographystyle{splncs04}
\bibliography{publication_ICPR}

\end{document}